\documentclass[twocolumn,pre,showpacs]{revtex4}
\usepackage{amssymb, amsmath, graphicx}
\def\be{\begin{equation}}
\def\ee{\end{equation}}
\begin{document}
\title{Cyclic Topology in Complex Networks}
\author{Hyun-Joo \surname{Kim}$^{1,2}$}\email{hjkim21@knue.ac.kr}
\author{Jin Min \surname{Kim}$^{2,3}$}\email{jmkim@physics.ssu.ac.kr}
\affiliation{
$^1$ Department of Physics Education, Korea National University of Education,
Chungbuk 363-791, Korea\\
$^2$ Department of Physics and CAMDRC, Soongsil University, Seoul 156-743, Korea\\
$^3$ Asia Pacific Center for Theoretical  Physics, POSTEC, Pohang, 790-784, Korea }

\received{\today}

\begin{abstract} 

We propose a cyclic coefficient $R$ which represents the cyclic
 characteristics
 of  complex networks.  If the network forms a perfect tree-like
structure then $R$ becomes zero. The larger value of $R$
 represents that the network is more
cyclic. We measure the cyclic coefficients and the distributions of the
local cyclic coefficient for both various real networks and the representative
network models and  characterize the cyclic structures of them.

\end{abstract}

\maketitle

During a recent few years, complex networks have received
considerable attentions \cite{review}. They appear in
a variety of system such as biological
\cite{metabolic,protein,bio}, social
\cite{actor,coauthor1,coauthor2,sex}, informational
\cite{internet,www}, and  economic \cite{stock} systems. Such
complex networks are characterized by some topological and
geometrical properties such as small world, high degree of
clustering, and scale-free topology. The small-world property
denotes that the average path length $L$ which is the average
shortest path length between vertex pairs in a network, is small.
It grows logarithmically with the network size $N$. The
clustering structure in a network is measured by the clustering
coefficient $C$ which is defined as the fraction of pairs between the
neighbors of a vertex that are the neighbors  of each other. The
high degree of clustering  indicates that if vertices A and B are
linked to vertex C then A and B are also likely to be linked to each
other. These two properties were realized by small-world network
(SWN) model \cite{smallworld} in which randomly selected vertex
pairs are linked by short-cuts. The scale-free (SF) topology
reflects that the degree distribution $P(k)$ follows a power law,
$P(k) \sim k^{-\gamma}$, where degree $k$ is the number of edges
incident upon a given vertex and $\gamma$ is the degree exponent.
An evolving model introduced by Barabasi-Albert (BA) \cite{ba}
well illustrated the SF property. Such network is called
the SF network.

Recently, many efforts have been done to elucidate
the structural properties of complex networks. The hierarchical
structure appears in some real networks and has been clarified
by power-law behavior of the clustering coefficient $C(k)$ as a
function of the degree $k$ \cite{ck1,ck2,ck3,ck4,ck5,ck6}. This
indicates that the networks are fundamentally modular and it is
the origin of the high degree of clustering of complex networks.
Also, it was recently found that many real networks include
statistically significant subnetworks, so-called motifs, in their
structures \cite{motif1,motif2,motif3}.

Especially, the recent studies for the topological properties of
complex networks have attracted much attention to the loop (cyclic)
structure. The presence of loops has some effect on the delivery of
information, transport process, epidemic spreading behavior
\cite{es}, and etc. With respect to a tree-like topology, loops
provide more paths along which the information or virus can
propagate. A cycle of order $k$ can be defined as a closed loop of
$k$ edges. That is, graphically a triangle is a cycle
of order 3, while a rectangular is a cycle of order 4. A tree
which can not form a closed loop can be regarded as a cycle of
infinite order. The clustering coefficient counts the triangle
structure only. Meanwhile there are many cycles of
higher order which is larger than 3 in complex networks so that
it is necessary to
investigate the cycles of higher order for the characterization of
the cyclic structure. Some previous studies \cite{lbc,lny,lcpv,lphy} in which
cycles of order 4 or 5 were considered, are good trials to explain
the loop structure of higher order.

In this paper, we survey the cyclic topology in complex networks
by introducing a new quantity $R$ which characterizes the degree of
circulation in the systems. We consider the cycles of all order
starting from three up to infinity to define the quantity $R$. By
monitoring the values of $R$ and its distribution, the cyclic topology of
the networks is analyzed for both several real
networks from technological to social systems and the network
models such as SWN and BA models. 

\begin{figure}
\includegraphics[width=7cm]{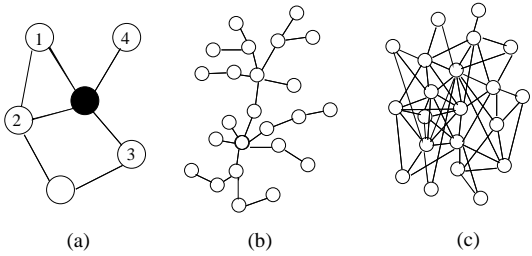}
\caption{ (a) A typical example for the cyclic coefficient. The
cyclic coefficient of filled circle is obtained by $r_\bullet =
0.13$. The two sample networks with same network size $N = 25$ and
different cyclic coefficient are shown in (b) and (c) where R = 0
and R = 0.29, respectively. } \label{loop}
\end{figure}

We introduce a new quantity $R$ to
measure how cyclic is a network and call it the cyclic
coefficient. At first, the local cyclic coefficient $r_i$ for a
vertex $i$ is defined as the average of the reciprocal of the size
of loops which are formed by a vertex $i$ and its two neighboring
vertices, i.e., \be
 r_i = \frac{2}{k_i (k_i -1)} \sum_{<lm>} \frac{1}{S_{lm} ^i}
\ee
where $k_i$ is the degree of a vertex $i$ and $<lm>$ is all the
pairs of the neighbors  of the vertex $i$. $S_{lm} ^i$ is the  smallest size of the
closed path that pass through a vertex $i$ and its two neighbor
vertices $l$ and $m$. If vertices $l$ and $m$ is directly connected to each
other then the vertices $i$, $l$, and $m$ form a triangle. It is
a cycle of order 3
and $S_{lm} ^i$ has a value three which is the smallest value of
$S$.  If there does not exist any paths that connect vertices $l$
and $m$ except for the path through the vertex $i$, then the vertices $i$, $l$,
and $m$ form a tree. In this case, there does not exist any loop
pathing through the three vertices $i$, $l$,
and $m$.  It is the cycle of infinite order and the value of $S_{lm} ^i$
becomes infinity. For an example as shown in Fig. \ref{loop} (a)
the local cyclic coefficient $r_{\bullet}$ of a vertex $\bullet$ is
given as $r_{\bullet} =  0.13$ with  $S_{12} ^{\bullet}= 3$,
$S_{23} ^{\bullet}= 4$, $S_{13} ^{\bullet}= 5$, and
$S_{14} ^{\bullet}= S_{24} ^{\bullet}= S_{34} ^{\bullet}= \infty$.

The cyclic coefficient $R$ is the average of $r_i$ over all the vertices,
$R$ $ = \langle r_i \rangle$ which has a value
between zero and $1/3$. $R$=0 means that a network has a perfect
tree-like structure in which no loops is formed. Meanwhile if
all the neighbor pairs are connected to each other i.e.,  the
clustering coefficient
becomes
$C=1$, and  $R$=1/3. Figure \ref{loop} (b) and (c) show two examples with $R$=0
and $R$=0.29 for $N = 25$, respectively. Thus the larger is the cyclic
coefficient $R$, the more cyclic is the architecture of the network. The
cyclic coefficient $R$ could be a good quantity to identify the
degree of circulation in a complex network.

In order to characterize the cyclic topology in real networks we have
measured the cyclic coefficient $R$ for several real networks \cite{real}
appearing in biological, technological, and social systems.
In the measurement, we excluded the isolated vertices
and focused on the entirely connected part of the network.

\begin{figure}
\includegraphics[width=7cm]{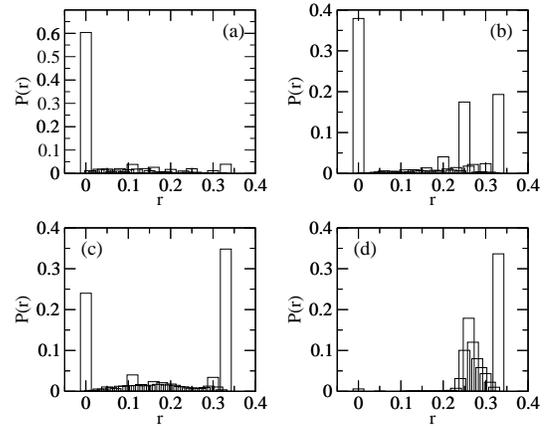}
\caption{
The probability distribution of local cyclic coefficient for four real networks,
(a) protein network, (b) Internet network, (c) math co-authorship network,
and (d) movie actor collaboration network.
}
\label{sdreal}
\end{figure}

First, we  consider the protein network \cite{protein} which
is composed of 1458 proteins. It has 1948 identified direct physical
interactions. The proteins and the direct interactions are
considered as vertices and edges, respectively. Figure
\ref{sdreal} (a) shows the histogram of the distribution $P(r)$ of
the local cyclic coefficient. About 60\% of the total
vertices have $r=0$ and $P(r)$ has small value for all the range  $0<r
\le 1/3$, resulting in small value $R$ $\approx$
0.06, which indicates that there are very little loops and the
network constitutes a tree-like structure. Thus a tree-like
topology of the protein network pictured in the reference
\cite{protein} is well quantified by our cyclic coefficient.

Second, the physical internet network \cite{internet} at the
inter-domain (Autonomous System(AS)) level is considered.
Each domain, composed of hundreds of routers and computers, acts
as a vertex and an edge is drawn between two domains if there is
at least one route that connects them. The network at the AS
level, as of 15th September 1999 is composed of both 5746 ASs and 11017
edges.  $R
\approx 0.16$ is obtained in this network.  From the distribution of the local cyclic
coefficient $r$ (Fig. \ref{sdreal} (b)), we found that the most
vertices have a value among $r=0$, $r=0.25$, and $r=1/3$. That is,
the vertices with a tree structure are dominant ($r=0$)  and the most of
the rest form loops of small size (3 or 4).

Third, we  consider the network of scientific collaborations
in the field of mathematics published in the period 1991-1998
\cite{coauthor1}, in which the vertices are the scientists. They
are connected if they write a paper together.  The total number of
vertices and edges are 57516 and 143778 respectively. The value of
cyclic coefficient is  $R \approx 0.19$. Figure
\ref{sdreal} (c) shows the probability distribution of $r$. It
has the first peak at $r=0.33$,
which indicates that cycles of order 3 dominate in the networks. 
distinguishing from the protein and internet networks where the
tree-like structure dominates.

Finally, the movie actor collaboration network \cite{actor} is
constituted of 9865 vertices and 273412 edges. The
vertices are the actors and  two vertices are connected if the
corresponding actors have acted in the same movie together. Figure
\ref{sdreal} (d) shows that the distribution of $r$ has a maximum
value at $r=0.33$, which reflects the high degree of clustering in
social networks. Meanwhile the vertices with $r=0$  almost do not
exist in contrast to the case of the other networks. This explains
that the movie actor
network is more cyclic with large value of $R=0.29$. 

From the results of the above four examples, 
we have found that both the well clustered parts and non clusterd parts
coexist in the real networks especially in the math co-authorship network.
The probability distribution $P(r)$ is not uniform. Instead, there are
a few peaks at  certain values of $r$ such as $r=0$ or $r=1/3$.    
It means that most of the vertices have triangle structure or
tree structure with the neighbor vertices. That is, the neighbor vertices 
in the well clustered parts have high connections each other while 
in the other parts there are no clustering at all.
Thus by measuring the distribution of local cyclic coefficient we can
understand the details of cyclic structure in the complex networks.

\begin{figure}
\includegraphics[width=7cm]{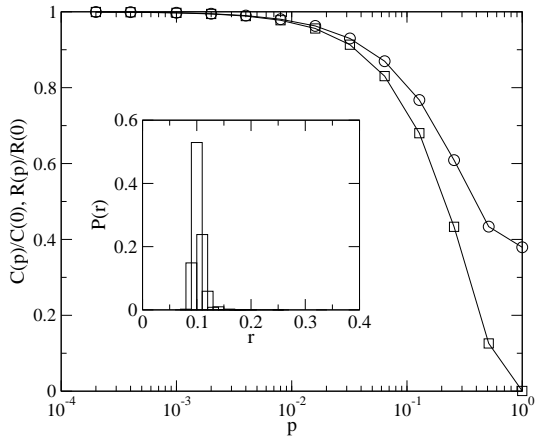}
\caption{ The cyclic coefficient $R(p)$ (circle) and the clustering coefficient
$C(p)$(square) for the SW network model. The data is normalized by the
$R(0)$ and $C(0)$ which are 0.283 and 0.5 respectively for a regular network.
The distribution of local cyclic coefficient for the random network ($p=1$)
is shown in the inset.} 
\label{swn}
\end{figure}
We have considered the cyclic coefficient $R$ for two representative models of complex
networks, the SWN \cite{smallworld} and the BA model \cite{ba}.
The algorithm of the SWN model is the following:
Consider a one-dimensional lattice of $N$ vertices with periodic boundary
conditions, i.e., a ring and connect each vertex to its first $m$ neighbors. The
small-world model is then created by randomly rewiring each edge of the lattice
with probability $p$, moving one end of that edge to a new location chosen
randomly from the lattice, except that self-connections and duplicate
edges are created.
This rewiring process introduces $pNm/2$ shortcuts which
connect vertices that are in long-range and by varying $p$ the transition between
a regular lattice ($p=0$) and random network ($p=1$) \cite{rg} can be shown.

Figure \ref{swn} shows the plot of the normalized clustering coefficient
$C(p)/C(0)$ and the cyclic coefficient $R(p)/R(0)$ as a function of
the rewiring probability $p$ with the
network size $N = 10000$ and $m = 4$. The clustering coefficient stays almost
unchanged for $p<0.01$   and drops to almost zero at $p=1$. It is the
characteristics of SWN with the high degree of clustering for $p<0.01$.
The cyclic coefficient $R(p)$ also keeps up the value of $R(0)$ for
$p < 0.01$  while decreases  to a finite value at $p=1$. 
The finite  value of $R(1)$  comes from the contribution of the loops
for all orders. 
The inset of Fig. \ref{swn} shows the properties of the random network ($p=1$).
As shown in the inset, the cyclic distribution 
has a peak at $r=0.11$ with $R \approx 0.11$. It is interesting that
$P(r)$ is almost zero for both $r=0$ and $r=1/3$.

\begin{figure}
\includegraphics[width=7cm]{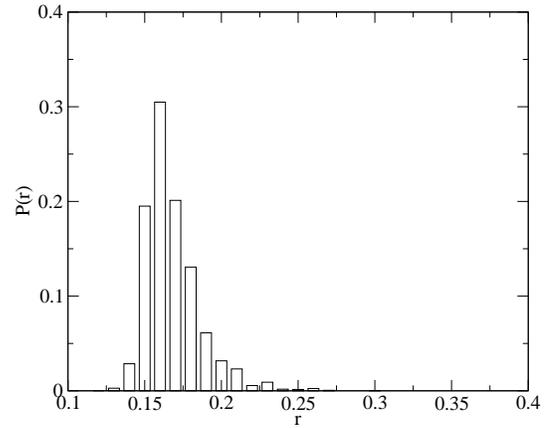}
\caption{ The distribution of local cyclic coefficient for the BA
model }
\label{ba}
\end{figure}

We also measure the cyclic coefficient for the BA model \cite{ba}.
The  BA model is carried out
as the following: Start with a small number $N_0$ of vertices
and no edges. At every time-step, a new vertex with $m$ ($<= N_0$)
edges is added where the $m$ edges link the new vertex to $m$
different vertices already present in the system. The vertices to
which the new vertex is connected are chosen with the preferential
attachment rule in which the probability $\Pi$ for a vertex $i$ to
be connected with a new vertex depends on the degree $k_i$ of the
vertex $i$, such that $\Pi (k_i) = {k_i}/{\sum_j k_j}$. 
We have obtained $ R \approx 0.17$ with the network size $N=10000$
for the BA model.  As shown in Fig. \ref{ba}, the distribution of the local
cyclic coefficient in the BA model shows a poisson-like shape having 
a peak at $r=0.16$. It represents the random nature of the
local circulation in the  BA networks. 
However, in the real networks given above  the cyclic distributions do not
follow the poisson-like shape and  have a peak at $r=0$ or $r=1/3$
. There is almost no mechanism to form a triangle
or tree structure in the BA model, in contrast  to the case of the real networks. 

The  network size $N$,
mean degree $\langle k \rangle$, clustering coefficient $C$, cyclic coefficient $R$,
the probability distribution with $r=0$ $P(0)$ (tree structure), and
 with $r=1/3$ $P(1/3)$ (cyclic structure of loops with  length three)
are summerized  in  Table \ref{tb} for the considered networks. 

\begin{table*}
\caption{\label{tb}
For both four real networks and two network model, we summarized  the
 various data of 
 network size $N$,
mean degree $\langle k \rangle$, clustering coefficient $C$, cyclic coefficient $R$,
the probability distribution with $r=0$ $P(0)$ (tree structure), and
 with $r=1/3$ $P(1/3)$ (cyclic structure of loops with  length three).
}
\begin{ruledtabular}
\begin{tabular}{lrccccc}
Network & N & $\langle$ k $\rangle$ & C & R & $P(0)$ & $P(1/3)$ \\
\hline
protein interactions & 1458 & 2.67 & 0.07 & 0.06 & 0.60 & 0.04 \\
Internet & 5746  & 3.83 & 0.24 & 0.16 & 0.38 & 0.19 \\
math co-authorship & 57516 & 5.00 & 0.48 & 0.19 & 0.24 & 0.35 \\
movie actor collaborations  & 9853 & 54.95 & 0.58 & 0.29 & 0.01 & 0.34 \\
\hline
random network ($p=1$) & 10000 & 4 & 0.0003 &0.11 & 0 & 0 \\
BA network & 10000 &  6 & 0.006 & 0.17 & 0 & 0  \\
\end{tabular}
\end{ruledtabular}
\end{table*}

In conclusion, we have evaluated the degree of circulation in
complex networks by introducing the cyclic coefficient $R$. It
inclueds  the total  effect of all the sizes of the loops.
  If the network forms a perfect
tree-like structure, $R$ becomes zero. The value of cyclic coefficient is in between zero
and 1/3.  The larger is the cyclic coefficient, the
more cyclic becomes a network. We measured the cyclic coefficients
for various real complex networks and the representative network
models. For the protein network of biological system the cyclic
coefficient is small which reflects that the protein network is
tree-like, while for the movie actor collaboration network of
social system we found that the cyclic coefficient is large and
its structure is more cyclic. Also by measuring the probability
distribution of the
local cyclic coefficient, we could classify the details of the cyclic
structure of the complex networks.  
Thus the cyclic coefficient and its
distribution help us to understand the cyclic structures of the complex
networks. It is interesting to keep in surveying the cyclic coefficient
for other various complex networks. 

This work was supported by Korea Research Foundation
Grant (KRF-2003-015-C00003).

\end{document}